\newcommand{\vect}[1]{\left(\begin{array}{c} #1 \end{array}\right)}
\newcommand{\matwo}[1]{\left(\begin{array}{cc} #1 \end{array}\right)}

\documentclass[aps,twocolumn,prl]{revtex4}
\usepackage{graphicx}
\usepackage[]{epsfig}
\begin{document}

\draft
\title{Using time reversal symmetry for sensitive incoherent matter-wave Sagnac interferometry}
\author{Y. Japha, O. Arzouan, Y. Avishai and R. Folman}
\date{\today}
\address{Department of Physics,
Ben-Gurion University, Be'er-Sheva 84105, Israel}

\begin{abstract}

We present a theory of the transmission of incoherent guided matter-waves through Sagnac interferometers.
Interferometer configurations with only one input and one output port
have a property  similar to  the phase rigidity observed in the transmission through
Aharonov-Bohm
interferometers in coherent mesoscopic electronics. This property is connected to the
existence of counterpropagating paths of equal length and enables the
operation of such matter-wave interferometers with incoherent sources. High finesse
interferometers of this kind have a rotation sensitivity inversely proportional
to the square root of the finesse.

\end{abstract}

\maketitle

The Sagnac effect, occurring in a wave propagating through a
closed rotating ring, induces a phase shift proportional to the
angular frequency $\Omega$ of this rotation and the area $A$ of
the ring \cite{Sagnac_review}. For light waves with frequency
$\omega$ this phase shift is $\phi_{\rm light}=(2\omega
A/c^2)\Omega$, where $c$ is the speed of light. For de-Broglie
waves of non-relativistic particles with mass $m$, the phase shift
is $\phi_{\rm matter}=(2mA/\hbar)\Omega=(mc^2/\hbar
\omega)\phi_{\rm light}$ \cite{Sagnac_neutrons,Sagnac_electrons},
so that the rotation sensitivity of a matter-wave interferometer
('ifm') is potentially better by a factor of $\sim 10^{11}$. First
experimental attempts demonstrated rotation sensitivities
comparable to or even better than those of optical Sagnac ifm's
\cite{Sagnac_Ca, Sagnac_Na,Kasevich}. However, these "one-pass"
ifm's are limited by their small effective area and the relatively
low flux available from coherent matter wave sources. Light ifm's
with "multi-pass" configurations ("high finesse"), such as ring laser gyros,  
may be used to increase the effective area and achieve good rotation sensitivity.
Recently waveguide ring structures for cold atoms were demonstrated
\cite{StorageRing,RadioRing}, opening the door for the development
of "multi-pass" guided atomic Sagnac interferometry. However, the
small de-Broglie wavelength and short coherence length (of the order of
1$\mu m$) of matter waves makes these ifm's very sensitive to imperfections in the
guiding potential, leading to effective path length fluctuations.

In this Letter we use the analogy \cite{Sakurai} between the
Sagnac effect for massive neutral particles and the Aharonov-Bohm
(AB) effect in coherent electron transmission through mesoscopic
rings \cite{Imry}. AB ifm's with only two ports connected to the
ring show the effect of "phase rigidity" of the transmission
pattern as a function of the magnetic flux $\Phi$, when an
effective path length difference between the ifm arms is
introduced \cite{PhaseRigidity,Aharony}. The locking of the
transmission pattern at $\Phi=0$ stems from the time reversal
invariance and current conservation properties of any system with
two ports, which implies that transmissivity is invariant to
magnetic field inversion $\Phi\rightarrow -\Phi$. We show that
this property, when applied to matter wave Sagnac ifm's, leads to
their robustness under effective path length differences and
enables their operation in a high finesse configuration and with
incoherent sources, which are available with higher particle flux.
Ifm's of this kind will enable increasing the rotation sensitivity
and their miniaturization onto atom chips \cite{folman} while
maintaining the required sensitivity.

In general, a Sagnac ifm consists of a loop and one or more
junctions, each consisting of one or more beam splitters (BS)
connecting the loop to input and output channels. Here we consider
a one dimensional (1d) model in which the particles are guided in a single
transverse mode of the waveguide. A linear junction with $n$ ports is
represented by a $n\times n$ unitary scattering matrix $S$
connecting the output amplitudes to the input amplitudes at the
ports. As demonstrated in fig.1 we denote the indices of the ports
of the input junction connected to the ifm loop by $\alpha$ and $\beta$
and the ifm input port by $i$. The corresponding input and output
amplitudes at the internal ports $\alpha$ and $\beta$ will be
denoted by $a_{\pm}$ and $b_{\pm}$ respectively and the ifm input
amplitude by $a_{in}$ (normalized to $a_{in}=1$).
The relation between the amplitudes is then given by
\begin{equation}
\vect{a_+ \\ b_+}= \vect{S_{\alpha i} \\ S_{\beta i}}a_{in}+\matwo{S_{\alpha\alpha} & S_{\alpha\beta} \\
S_{\beta\alpha} & S_{\beta\beta}}\vect{a_- \\ b_-}.
\label{apbp}
\end{equation}
If the system is linear, then the amplitudes
$a_-$,$b_-$ are related to the amplitudes $a_+$,$b_+$ by
\begin{equation}
\label{relT}
\vect{a_- \\ b_-}=S_L\vect{a_+ \\ b_+}
\end{equation}
where the 2$\times$2 $S$-matrix $S_L(k,\Omega)$ describes the
transmission through the loop, thereby containing terms of the
form $e^{i(kL\pm\phi)}$, $\hbar k$ being the longitudinal momentum
of the particles, $L$ the circumference of the loop and $\phi$ the
rotational phase shift. $S_L$ is in general non-unitary when
particles can leave the loop from another junction. Time reversal
symmetry (Onsager relations \cite{Onsager}) implies that
$S_{\alpha\beta}=S_{\beta\alpha}$ and that $S_L$ has  the form
$S_L=\matwo{a & be^{i\phi} \\ be^{-i\phi} & c}$, where $a,b$ and
$c$ are functions of $k$. By combining  eqs.~(\ref{apbp})
and~(\ref{relT})  we find the solution
\begin{equation}
\vect{a_+ \\ b_+}=\frac{1}{I-T} \vect{S_{\alpha i} \\
S_{\beta i}}a_{in}= \sum_{n=0}^{\infty}T^n
\vect{S_{\alpha i} \\ S_{\beta i}} \label{GenSol}
\end{equation}
where I is the 2$\times$2 unity matrix and  $T=\tilde{S}S_L$,
$\tilde{S}$ being the sub-matrix of $S$ appearing in
eq.~(\ref{apbp}).  The output of the ifm is obtained when
propagating the amplitudes $a_+$,$b_+$ through the ifm arms and
transmitting them into the output port through the output
junction. When no additional imperfections exist in the arms, the
output amplitude is given by
\begin{equation}
a_{out}=S'_{o\alpha'}e^{i(k-\phi/L)L_{\alpha}}a_+ + S'_{o\beta'}e^{i(k+\phi/L)L_{\beta}}b_+,
\label{a_out}
\end{equation}
where $S'$ is the scattering matrix of the output junction and
$L_{\alpha},L_{\beta}$ are the lengths of the corresponding arms,
such that $L=L_{\alpha}+L_{\beta}$.

Time reversal symmetry, which determines the symmetry properties
of the scattering matrices $\tilde{S}$ and $S_L$ discussed above
yields the following general form of the transmission through the
ifm \cite{Aharony}
\begin{equation}
P(k,\phi)=|a_{out}|^2=\frac{B+C\cos(\phi+\beta)}{1+D\cos\phi+H\cos^2\phi},
\label{genform}
\end{equation}
where $B,C,D,H$ and $\beta$ are functions of $k$. When the ifm has
only one input and one output port, time-reversal symmetry and
current conservation imply that $P(\phi)=P(-\phi)$ so that
$\beta=0$ or $\beta=\pi$ ("phase rigidity"). In what follows we
show that while in an ifm having more than two open ports $\beta$
may be strongly $k$-dependent, such that an integration over a
wide momentum bandwidth $\Delta k$ washes out the $\phi$
dependence of $P(\phi)$, this dependence is conserved in an ifm
having only 2 open ports and fixed $\beta$, enabling wide momentum
bandwidth operation.

\begin{figure}
\includegraphics[width=7cm]{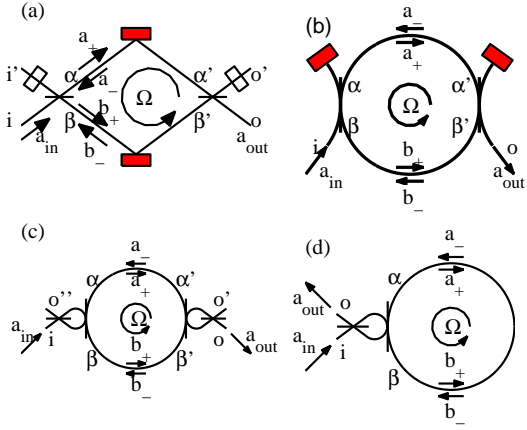}
\caption{Geometries of guided matter-wave Sagnac interferometers:
(a) MZ. Open squares mean controllable reflectivity. When fully
reflecting we name the ifm: closed MZ. (b) 2-port loop. Closed
squares are fully reflecting mirrors. (c) 4-port loop. (d) single
junction loop. In all configurations horizontal lines mean 50-50\%
BSs while vertical lines mean BSs with a transmission amplitude
$i$t.}
\end{figure}

Let us first examine a simple Sagnac ifm which does not satisfy
the condition for "phase-rigidity". The Mach-Zehnder (MZ) ifm
shown in fig.1a is analogous to the ifm implemented in
\cite{Kasevich}.
It contains two 50-50\% junctions where an incident particle may
be either reflected with amplitude
$S_{\beta,i}=S_{\alpha,i'}=1/\sqrt{2}$ or transmitted with
amplitude $S_{\alpha,i}=S_{\beta,i'}=i/\sqrt{2}$.  This is a
one-pass ifm, where no reflections occur from the loop ports
$\alpha$ and $\beta$ back into the loop ($\tilde{S}=S_L=T=0$), so
that the transmission probability at the output port $o$ is
$P(k,\phi)=\sin^2[(\phi-k\delta L)/2]$, where $\delta
L=L_{\alpha}-L_{\beta}$ is the length difference between the ifm
arms. $P(k,\phi)$ can be then written in the form (\ref{genform})
with $B=-C=1/2$, $D=H=0$ and $\beta=k\delta L$. If $\delta L\neq
0$ and the input flux is described by a Gaussian spectral
distribution $G(k)$ with bandwidth $\Delta k$ around $k=\bar{k}$,
then the time-averaged transmission probability $P(\Omega)=\int dk
G(k) P(k,\phi(\Omega))$ becomes
\begin{equation}
P(\Omega)=\frac{1}{2}\left[1-\eta \cos\left(\frac{2mA}{\hbar}\Omega-\bar{k}\delta L\right)\right]
\label{PMZ}
\end{equation}
where the interference visibility $\eta=e^{-\Delta k^2 \delta
L^2/2}$ decreases with momentum bandwidth and path difference. If,
in addition, the atomic beam is not perfectly collimated so that
parts of it take paths surrounding different effective area with
uncertainty $\delta A$, then $\eta$ is multiplied by a factor
$\exp[-(2m/\hbar)^2\delta A^2 \Omega^2/2]= \exp[-(\delta A^2/A^2)
\phi^2/2]$. Both suppression factors have been observed
experimentally, as demonstrated in fig.3 of ref.~\cite{Kasevich}.

In contrast, we now study a closed loop ifm obtained by closing
the ports $i'$ and $o'$ of the MZ with mirrors placed in front of
them (fig.1a), so that each junction now has only 3 open ports. A
particle incident from one of the loop arms on such a junction may
be reflected back into the loop with amplitude
$S_{ij}=e^{i\varphi_{ij}}/2$, where each phase $\varphi_{ij}$
($i,j= \alpha,\beta$) consist of a contribution $\theta$ upon
reflection at the mirror and an additional $\pi/2$ phase for each
transmission through the BS. We then have $S_{\alpha\alpha}=
-S_{\beta\beta}=e^{i\theta}/2$ and $S_{\alpha\beta}=
S_{\beta\alpha}=ie^{i\theta}/2$. The matrix $S_L$ has elements similar to
the elements of the $S$-matrix of the output junction,  $(S_L)_{ij}=S'_{ij}e^{i\eta_{ij}}$,
where $\eta_{ij}$ describe propagation through the arms: $\eta_{ij}=kL\pm\phi$ for
$i\neq j$ and $\eta_{jj}=2kL_j$. Following the above prescription, we obtain
the transmission probability as in~(\ref{genform}) with $\beta=0$,
such that $P(\phi)=P(-\phi)$. In order to study the properties of
the transmission when the source has a momentum bandwidth $\Delta
k$, it is insightful to realize, in view of eqs.~(\ref{GenSol})
and~(\ref{a_out}), that the output amplitude can be written as a
sum $a_{out}=\sum_{n=0}^{\infty}a_n e^{i(n+1/2)kL}$, where the
amplitudes $a_n$ are functions of $k$ with terms $e^{ik\delta L}$,
which are slowly oscillating relative to the fast oscillation of
$e^{inkL}$, if we assume small path length differences $\delta
L\ll L$. In the transmission probability $P(k,\phi)=|a_{out}|^2$
these fast oscillations will appear in cross terms $a_n^*a_{n'}$
with $n\neq n'$ describing interference between trajectories with
different number of passes through the loop in either direction.
In a realistic situation where the coherence length of the
matter-wave source is much smaller than $L$  ($\Delta k \gg
2\pi/L$), these interference terms will be eliminated.  We
describe this elimination by defining a slowly varying
time-averaged transmission probability integrated over a period
$2\pi/L$ of the fast oscillations
\begin{equation}
\bar{P}(k,\phi)\equiv
\left(\frac{L}{2\pi}\right)\int_{k-\pi/L}^{k+\pi/L}dk'
|a_{out}(k',\phi)|^2 \approx \sum_{n=0}^{\infty}
|a_n|^2, \label{nExpan}
\end{equation}
which describes the transmission of a quasi-monochromatic flow ($\Delta k\delta L\ll 1$), where
the coherence length is large enough relative to effective path length differences between
trajectories  with the same number of passes. In the incoherent
limit where $\Delta k\delta L\gg 1$, integration of $P(k,\phi)$ over the
bandwidth  is equivalent to taking
the average of $\bar{P}$ over $0<k\delta L< 2\pi$. In this limit
only paths with exactly the same length may interfere.

For the closed MZ the transmission probability in the
quasi-monochromatic case is found to be
\begin{equation}
\bar{P}=\frac{1-1/2(\cos ^2 k\delta L+\cos^2\phi)}{1-1/4(\cos k\delta L-\cos\phi)^2},
\end{equation}
as shown in fig.2a for a few values of $k\delta L$ (thin curves).
In the incoherent limit (thick curve) the visibility is not
suppressed as in the simple MZ but stays fixed at $\eta\approx
0.16$. A more general explanation for this residual visibility is
interference between counterpropagating waves that follow trajectories with 
exactly equal length, as in white light interferometry. Change of the rotation frequency
$\Omega$ changes the relative phase  between these trajectories but they still interfere. 
In the closed MZ the existence of such pairs of interfering trajectories
is allowed by internal reflections, which in a system with a single
transverse mode  are an inevitable result of current conservation in a 
junction with only three ports.

The rotation sensitivity of a Sagnac ifm is the minimal rotation
frequency change $\delta\Omega_{min}$ that generates a
noticeable transmission change (beyond noise level). For
an average incident flux $F$ with a Poissonian
particle number distribution
\begin{equation}
\delta\Omega_{min}=\frac{\hbar}{2mA}\frac{\delta\phi_{min}}{\sqrt{F\tau}},\,\,
\delta\phi_{min}=\sqrt{P(\phi)}\left|\frac{\partial
P}{\partial \phi}\right|^{-1},
\label{sensitivity}
\end{equation}
where $\tau$ is the measurement integration time  and $\delta\phi_{min}$ the
dimensionless phase sensitivity per particle.
The best sensitivity is achieved approximately near points with
maximal derivative of the transmission. For ifm's with sinusoidal
transmission as the MZ, the best sensitivity is inversely
proportional to the visibility $\eta$.

In order to achieve better rotation sensitivity we now turn our
attention to ifm's with a high finesse loop, where the finesse (${\cal F}$) is
defined as the ratio between the periodicity of the spectral transmission $2\pi/L$ and 
the resonance bandwidth $(1-R)/\sqrt{R}L$, $R$ being the probability to stay in the loop 
for a full round-trip.  For a high finesse loop ($R\approx 1$), ${\cal F}$ is proportional to 
the average number of passes of a particle through the loop before exiting through a 
junction. The closed MZ ifm can be
converted into a high finesse ifm by replacing the BS of the
closed MZ with a "vertical" BS rotated by 90$^{\circ}$ relative to
the MZ BS (fig.1b). A particle incident on the "vertical" BS has a
reflection amplitude $r$  and a transmission amplitude
$it=i\sqrt{1-r^2}$, which is controllable, for example, with a
magnetic tunneling barrier as suggested in \cite{erika}. ${\cal
F}\gg 1$ requires that $r\approx 1$ and $t\ll 1$. Here the
transmission amplitude between the two arms $S_{\alpha\beta}=
S_{\beta\alpha}=r$, while back reflection through the mirror is
only permitted to arm $\beta\beta'$ with amplitude
$S_{\beta\beta}=-t^2e^{i\theta}$. An analysis similar to that of
the closed MZ yields the quasi-monochromatic transmission
probability shown in fig.2b for different values of $k\delta L$
(thin curves). The transmission has a symmetric dip at $\phi=0$,
whose depth and width depends on the value of $k\delta L$. 
Its width for a given value of $k\delta L$ is proportional to the spectral 
bandwidth $\Delta_k=(1-r^4)/r^2 L\approx 2t^2/rL$ of the loop, which is inversely proportional to its 
finesse ${\cal F} =(2\pi/L)/\Delta_k\approx \pi r/t^2$. In the
incoherent limit (thick curve), 
$\bar{P}(\phi)\approx P_0[1-\eta{\cal L}(\phi)]$, where
$P_0\approx t^2/r$, ${\cal L}(\phi)$ is a Lorenzian of FWHM $\Delta_{\phi}\sim
\Delta_k L$ and $\eta\approx 0.175$. Using eq.~(\ref{sensitivity}), the best sensitivity  
near $\phi\approx \pm\Delta_{\phi}/2$  is $\delta\phi_{min}\sim \Delta_{\phi}/\sqrt{P_0}/\eta
\approx \sqrt{2\Delta}/\eta_{\phi} \propto {\cal F}^{-1/2}$. The  transmission dip at 
$\phi=0$ is due to the fact that interference between pairs of equal length paths 
terminating at the output port is destructive when $\phi=0$. For comparison, 
we calculated the transmission of a similar
high finesse ifm with additional ports using a combination of a
50-50\% BS and a "vertical" BS at each junction (fig.1c). The
quasi-monochromatic transmission of this ifm (thin curves in
fig.2c) is not symmetric about $\phi=0$ and the visibility drops
to 0 in the incoherent limit (thick curve).

\begin{figure}
\includegraphics[width=7cm]{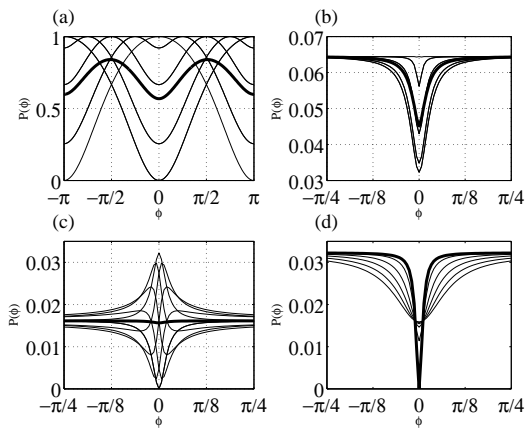}
\caption{Transmissivity of Sagnac ifm's as a function of rotational phase shift $\phi$
 (a) closed MZ.  (b) 2-port loop. (c) 4-port loop.  (d) single junction loop.
Thin curves in plots (a-c) represent the quasi monochromatic
transmission $\bar{P}(\phi)$ for few values of phase difference
between the two arms $k\delta L$, and in plot (d) the transmission
in the presence of a scattering center at $x=L/3$ with reflection
amplitudes from $r=0$ (thick curve) to $r=0.25$. Thick curves in (a-c)
represent the transmission in the incoherent limit while all curves in (d)
apply both in the quasi monochromatic and the incoherent limit.
For all four models $t=0.25$. As can be observed, the
phase locking in (c) suppresses the asymmetric functions appearing
in (b), giving rise to less smearing of the interference pattern.
The final dip in the incoherent signal of (c) may be explained as
interference only between exact same length trajectories, which
due to the $\pi/2$ phase difference between the r and t amplitudes, give rise to
destructive interference.}
\end{figure}

Finally we utilize the formalism developed here to analyze  a
simple ifm where each trajectory has a counterpropagating counterpart of the 
same length.  The
ifm configuration in fig.1d contains only one junction for input
and output. Its output amplitude is obtained by substituting
$L_{\alpha}=L_{\beta}=L$ and $S'=S$ in eq.~(\ref{a_out}). The
matrix $T$ for this ifm is diagonal, corresponding to no coupling
between counterpropagating modes, but on the other hand, all
trajectories of the same order $n$ have exactly the same length,
giving rise to destructive interference at $\phi=0$. The slowly
varying transmission probability (\ref{nExpan}) is then a sum over
$k$ independent contributions $t^4 r^{2n} \sin^2[(n+1)\phi]$
resulting in
\begin{eqnarray}
\bar{P}(\phi)&=& \frac{t^2}{1+r^2}\left[1-\frac{\cos^2\phi}{1+4(r/t^2)^2\sin^2\phi}\right]
\label{PSJ}
\end{eqnarray}
with a Lorenzian dip at $\phi=0$ of FWHM  $\Delta_{\phi}=t^2/r$, full
visibility ($\eta=1$) and an asymptotic transmission $P_0\approx
t^2/2r$, as shown in fig.2d (thick curve). The best sensitivity at
$\phi\rightarrow 0$ is $\delta\phi_{min}\approx t/(2\sqrt{2r})=\sqrt{\Delta_{\phi}/2}/2$,
which scales as the inverse square root of the finesse ${\cal F}=2\pi r/t^2$. The
transmission visibility is not affected by changes in the
effective path length, but may be affected by internal reflections
from asymmetrically located imperfections in the guiding
potential. This affect may be calculated from
eq.~(\ref{GenSol}) when the effect of scattering centers is
included in the loop $S_L$ matrix. As an example, we
present in fig.2d the incoherent transmission of the ifm when a
single scattering center with reflectivity amplitude $0\leq r\leq
0.25$ is placed at $x=L/3$. The sensitivity is then degraded both
by visibility reduction and increase of the Lorenzian width, while
the minimum transmission point stays at $\phi=0$. The effect of a
non-dispersive scattering center (which does not change the
effective path length) is similar for other high finesse ifm's
discussed above, regardless of their phase rigidity property.

To get an estimate of the achievable sensitivity of a Sagnac ifm
on an atom chip, we assume that an atom waveguide ring of radius
1cm is formed near the chip by magnetic field gradients of the
order $|{\bf \nabla}{\bf B}|\sim$~G/$\mu$m, which may be generated
by wires on the chip, about 10$\mu$m from the surface. The
centrifugal force $mv^2/r$ of the circulating atoms must not
exceed $\mu_B|{\bf \nabla}{\bf B}|$. This limits the maximum
velocity of the atoms to $v\sim 10$m/sec. An atomic trap lifetime
of about 10s permits up to 1000-2000 rotations, corresponding to a
tunneling amplitude $t\sim 0.035$ at the BS. For the single
junction loop we obtain $\delta\phi_{min}\sim 0.013$. If we assume
a flux of $10^9$ atoms per second (e.g. from a 2d MOT
\cite{2DMOT}) we obtain a sensitivity of $\delta\Omega\sim 5\cdot
10^{-13}rad/sec/\sqrt{Hz}$ - about three orders of magnitude
better than the best value published to date \cite{best_result}.

The 1d waveguide model put forward above can be easily extended to
a multi-mode model where a waveguide supports $N>1$ transverse
modes. We then have to replace the amplitudes
$a_{in},a_{out},a_{\pm}$ and $b_{\pm}$ with $N$ component vectors
and the matrices $\tilde{S},S_L$ and $T$ with $2N\times 2N$
matrices, which may also couple between the modes. 
The application of the suggestions in this Letter to real systems
requires a more comprehensive study of the multi-mode case
as well as other important effects, such as
dispersion of ifm components, Berry's phase in a magnetic ring
potential \cite{BerryPhase} and atom-atom collisions. The ideas
raised in this work concerning time reversal 
symmetry in a rotating quantum system with many possible paths are
closely related to the theory of weak localization and coherent back
scattering in mesoscopic electronic systems in the presence of a magnetic
field\cite{Imry}.



\begin{acknowledgments}
We thank Daniel Rohrlich, Ora Entin-Wohlman and Amnon Aharony for
useful discussions. R.F. would like to sincerely thank Yoseph
(Joe) Imry, for years of inspiration and specifically for his
insight into quantum mechanics and interferometry. We gratefully
acknowledge the support of the European Union FP6 'atomchip' (RTN)
collaboration, the German Federal Ministry of Education and
Research (BMBF-DIP project), the American-Israeli Foundation (BSF)
and the Israeli Science Foundation.
\end{acknowledgments}

\end{document}